\documentclass{aa}
\usepackage{graphicx}
\usepackage{graphics}
\usepackage{natbib}
\usepackage{txfonts}
\usepackage{float}
\usepackage{isotope}
\usepackage{soul} 
\usepackage[colorlinks=true,citecolor=blue,linkcolor=blue]{hyperref}
\bibpunct{(}{)}{;}{a}{}{,}
\begin{document} 

\title{Modelling the evolution of the Sun's open and total magnetic flux}
\author{N. A. Krivova\inst{1}
         \and
        S. K. Solanki\inst{1,2}
         \and
        B. Hofer\inst{1,3}
        \and
        C.-J. Wu\inst{1}
         \and
        I. G. Usoskin\inst{4}
         \and
        R. Cameron\inst{1}
}

\institute{Max-Planck-Institut f\"ur Sonnensystemforschung, Justus-von-Liebig-Weg 3, G{\"o}ttingen, Germany
  \and
  School of Space Research, Kyung Hee University, Yongin, Gyeonggi-Do,446-701, Republic of Korea
    \and
  Institut f\"ur Astrophysik,
  Georg-August-Universit\"at G\"ottingen,  Friedrich-Hund-Platz 1, 37077, G\"ottingen, Germany
  \and
  University of Oulu, Finland
}
\date{}

%

\abstract{
    Solar activity in all its varied manifestations is driven by the magnetic field. Particularly important for many purposes are two global quantities, the Sun's total and open magnetic flux, which can be computed from sunspot number records using models. Such sunspot-driven models, however, do not take into account the presence of magnetic flux during grand minima, such as the Maunder minimum. 
    Here we present a major update of a widely used simple model, which now takes into account the observation that the distribution of all magnetic features on the Sun follows a single power law. The exponent of the power law changes over the solar cycle. This allows for the emergence of small-scale magnetic flux even when no sunspots are present for multiple decades and leads to non-zero total and open magnetic flux also in the deepest grand minima, such as the Maunder minimum, thus overcoming a major shortcoming of the earlier models. 
    The results of the updated model compare well with the available observations and reconstructions of the solar total and open magnetic flux. This opens up the possibility of improved reconstructions of sunspot number from time series of cosmogenic isotope production rate.}

\keywords{Sun: activity - Sun: heliosphere - Sun: magnetic fields - Sun: photosphere - solar-terrestrial relations}

\maketitle

\section{Introduction}

The precise history of solar activity and its underlying magnetic field is of interest for a number of reasons. Firstly, records of solar activity and of the magnetic field pose an important constraint on models for the enhancement and the evolution of magnetic flux (mainly dynamo models). Secondly, such records are important for  understanding the history of the Sun's influence on the Earth (either through changes in its irradiance, or through space weather effects). Thirdly, a long record of solar activity is needed to understand how the Sun compares with other sun-like stars in its level of activity and variability \citep[e.g.,][]{Radick:2018,Reinhold:2020}.

Solar activity, in all its diverse manifestations, is driven by its magnetic field, so that knowledge of the history of solar activity implies knowledge of its magnetic field.
Two widely used quantities describing the global magnetic field of the Sun are the global open and unsigned total magnetic flux. They are, for example, used in heliospheric physics, for the reconstruction of solar irradiance, or as measures of solar activity when comparing with other stars. These quantities, being global in nature, can be reconstructed from more indirect proxies of solar activity and magnetism, such as the sunspot number and concentrations of the cosmogenic isotopes $^{14}$C or $^{10}$Be in terrestrial archives. 

A first model to compute the solar open flux from the sunspot number was developed by \citet{Solanki2000}, based on a simple differential equation describing the evolution of the open flux, $\phi_{\rm open}$. In spite of its simplicity, it successfully reproduced the empirically-reconstructed evolution of the open flux by \cite{Lockwood1999} and the $^{10}$Be concentration in ice cores \citep{Beer+al:1990}. This simple model was extended by \citet{Solanki2002} to cover also the total unsigned magnetic flux, $\phi_{\rm total}$, but now requiring the solution of a set of coupled differential equations to describe the evolution of ephemeral regions (ERs) besides that of active regions (ARs) and of the open flux. All three components of magnetic flux contribute to the evolution of the total magnetic flux.
ARs are the large bipolar structures that harbour sunspots at least part of the time, whereas ERs are smaller bipolar regions without sunspots.

The ability of this model to reproduce concentrations of cosmogenic isotopes turned out to be particularly useful \citep[e.g.,][]{usoskin_JGR_02}. Although far more sophisticated models are in the meantime available to compute not just global magnetic quantities, but also the underlying spatial distribution of the magnetic flux and the detailed input from individual emerging ARs etc., the very simplicity of this set of models allowed them to be inverted \citep[e.g.,][]{Lockwood:2003}, so that, e.g., sunspot number could be reconstructed from measured concentrations of cosmogenic isotopes \citep[e.g.,][]{Usoskin2003,Usoskin2004,Solanki2004,usoskin_AA_16,Wu18_composite}. 
 The model of \citet{Solanki2002} was further extended and combined with the successful SATIRE model \citep[Spectral And Total Irradiance REconstruction;][]{Fligge2000,Krivova2003,Krivova2011} to compute total solar irradiance over the last 400 years \citep{Krivova2007,Krivova2010}.
 \citet[][hereafter VS2010]{VieiraSolanki2010} have further refined the model by distinguishing between the short-lived and long-lived components of the open flux  
(\citealp{Ihksanov-Ivanov:1999,Cranmer:2002,Crooker+al:2002};  see \citealp{VieiraSolanki2010} for details), which led to an improved reconstruction of the open flux that displayed a better agreement with observations.
 This model, with some tuning, has been the basis for further reconstructions of solar spectral irradiance over the telescope era \citep{Krivova2010}, as well as sunspot number and TSI over the Holocene \citep{Vieira+al:2011,Wu18_composite,Wu18_ssi}.

One shortcoming of earlier versions of the model discussed above is that the open flux during a grand minimum, such as the Maunder minimum, i.e. during a long period essentially without sunspots, invariably drops to zero. This is because in this model, the emergence rate of the magnetic field on the solar surface is linearly linked to the sunspot number, so that  by design during a grand minimum no magnetic flux is allowed to emerge.
This leads to a zero open and total flux during the grand minima.
It has been shown, however, that signals of solar activity and variability were also present during the Maunder minimum 
\citep{Beer+al:1998,Fligge+al:1999,Usoskin2001,Miyahara:2004,Riley:2015}.
This was also confirmed  by modelling \citep{owens12} and points to a need for an improvement of the global total and open magnetic flux model. 
 
Furthermore, more recent solar observations provided new insights into the sources, emergence and evolution of the solar magnetic flux.
 Thus, \citet{Thornton-Parnell:2011} have combined observations from various sources and found that the
 emergence rate of bipolar magnetic regions with fluxes between $10^{16}$~Mx and $10^{23}$~Mx follows a single power law with a slope of $-2.69$.

Here we present a new, strongly revised version of the VS2010 model that builds on these recent solar observational results, replacing the direct proportionality of ERs and ARs by a more up-to-date approach. It does keep the original differential equations, however, so that it is not a completely independent model. As a natural outcome of the model, ERs keep emerging even during a grand minimum when there are no sunspots for multiple decades. This means that neither the open nor the total magnetic flux drop to zero at any time. 

The paper is structured as follows. The data used to constrain and test the new model are briefly introduced in Sect.~\ref{sec:data}.
We describe our model and highlight the changes relative to the older version of the model in Sect.~\ref{sec:model}. The results of the model are presented in Sect.~\ref{sec:resu}, while we summarise and discuss our findings in Sect.~\ref{sec:Summary}, where we also provide an outlook on future applications of the new model.

\section{Data}
\label{sec:data}

The model to be detailed in Sect.~\ref{sec:model} starts from a sunspot number time series and computes the total and open magnetic fluxes of the Sun therefrom.
We therefore require a sunspot series as input to the model.
To constrain the free parameters of the model, we compare its output to observations and independent data-based reconstructions of the total and open magnetic fluxes. 
Finally, to test the output of the model, we consider further independent time series of the reconstructed open magnetic flux.

As input to the model we use the following sunspot number data sets: (1) the international sunspot number v2.0, referred to hereafter as ISN2.0  \citep{clette16},
and (2) the group sunspot number, or GSN in short \citep{Hoyt:1998}.
The ISN2.0 data set was extended back to 1643 by adding the sunspot data during the Maunder minimum by \citet{Vaquero15} scaled up by the factor 1.67 to match the ISN2.0 definition.
Also in the GSN record, the values before 1710 were replaced by the data from \citet{Vaquero15}, without any scaling. 

To constrain the output of the model we make use of observations of the total magnetic flux \citep[see][]{Arge:2002,Wang:2005,Wenzler:2006}
derived from synoptic charts produced by the three solar observatories with the longest running regular magnetographic measurements: 
Wilcox Solar Observatory (WSO),
Mount Wilson solar Observatory (MWO), and National Solar Observatory at Kitt Peak (NSO/KP).
These data sets have already been used to constrain earlier versions of this model and we have used the same versions as employed by \citet{Krivova2007,Krivova2010} and \citet{Wu18_ssi}.
To constrain the free parameters of the model, we consider the average over at least two (MWO and WSO during 2002--2009) or all three (1976--2002) records for each Carrington rotation. 

Furthermore, to better constrain the free parameters of the model, we also use the empirical reconstruction of the open magnetic flux from the geomagnetic aa-index covering the period from 1845 to 2010 \citep{Lockwood14_geo}.

Finally, the quality of the computed open magnetic flux is tested by comparing it with two other independent data sets (without changing the free parameters of the model):
(1) a compilation of spacecraft-based in-situ measurements by \citet{Owens:2017} since 1998, and
(2)
a reconstruction by \citet{Wu18_composite}
from decadal INTCAL13 $^{14}$C data covering the Holocene prior to 1900 \citep{reimer13}.

\section{Model}
\label{sec:model}

\subsection{Magnetic flux emergence and evolution}
\label{sec:emerge}

Following the approach by \citet{Solanki2000,Solanki2002} and \citet{VieiraSolanki2010}, we describe the evolution of the solar total and open magnetic flux by a system of ordinary differential equations. However, instead of distinguishing between active regions (AR) and ephemeral regions (ER) as the sources of fresh magnetic flux at the solar surface, we distinguish between ARs and what  we call Small-Scale-Emergences (SSEs), i.e. all emergences with fluxes smaller than those in active regions. These therefore combine the flux emerging in the form of ephemeral regions and smaller magnetic bipoles all the way down to internetwork fields. The equations describing the evolution of the different (globally averaged)  components of the magnetic flux are:
\begin{equation}
\frac{{\rm d}\phi_{\rm AR}}{{\rm d}t} = \varepsilon_{\rm AR} - \frac{\phi_{\rm AR}}{\tau^{\rm 0}_{\rm AR}} - \frac{\phi_{\rm AR}}{\tau^{\rm s}_{\rm AR}} - \frac{\phi_{\rm AR}}{\tau^{\rm r}_{\rm AR}},
\label{eq:ode1}
\end{equation}
\begin{equation}
\frac{{\rm d}\phi_{\rm SSE}}{{\rm d}t} = \varepsilon_{\rm SSE} - \frac{\phi_{\rm SSE}}{\tau^{\rm 0}_{\rm SSE}} - \frac{\phi_{\rm SSE}}{\tau^{\rm s}_{\rm SSE}},
\label{eq:ode2}
\end{equation}
\begin{equation}
\frac{{\rm d}\phi^{\rm r}_{\rm open}}{{\rm d}t} = \frac{\phi_{\rm AR}}{\tau^{\rm r}_{\rm AR}} - \frac{\phi^{\rm r}_{\rm open}}{\tau^{\rm r}_{\rm open}},
\label{eq:ode3}
\end{equation}
\begin{equation}
\frac{{\rm d}\phi^{\rm s}_{\rm open}}{{\rm d}t} = \frac{\phi_{\rm AR}}{\tau^{\rm s}_{\rm AR}} + \frac{\phi_{\rm SSE}}{\tau^{\rm s}_{\rm SSE}} - \frac{\phi^{\rm s}_{\rm open}}{\tau^{\rm s}_{\rm open}},
\label{eq:ode4}
\end{equation}
\begin{equation}
\phi_{\rm open} = \phi^{\rm r}_{\rm open} + \phi^{\rm s}_{\rm open},
\label{eq:ode5}
\end{equation}
\begin{equation}
\phi_{\rm total} = \phi_{\rm AR} + \phi_{\rm SSE} + \phi_{\rm open}.
\label{eq:ode6}
\end{equation}
Here $\phi_{\rm AR}$, $\phi_{\rm SSE}$, $\phi_{\rm open}$ and $\phi_{\rm total}$ refer to the magnetic flux in ARs and SSEs, as well as the open, and total magnetic flux (all magnetic fluxes are global, i.e. referring to their unsigned sum over the entire solar surface).
The open flux is divided into rapidly ($\phi^{\rm r}_{\rm open}$) and slowly ($\phi^{\rm s}_{\rm open}$) evolving components.
$\tau^{\rm 0}_{\rm AR}$, $\tau^{\rm 0}_{\rm SSE}$, $\tau^{\rm r}_{\rm open}$, and
$\tau^{\rm s}_{\rm open}$ are the decay timescales of ARs, SSEs, rapid open flux, and slow open flux, respectively. 
The flux transfer timescales from ARs and SSEs to slow open flux are $\tau^{\rm s}_{\rm AR}$ and $\tau^{\rm s}_{\rm SSE}$, respectively, while the corresponding timescale for the transfer of  AR flux to the rapid open flux is $\tau^{\rm r}_{\rm AR}$. See \cite{VieiraSolanki2010} for a discussion on the distinction between rapidly and slowly evolving open magnetic flux.
 
In the original model, the emergence rate of ARs at a given time $t$, $\varepsilon_\mathrm{AR}(t)$, was linked linearly to the sunspot number, SN, at that time:
\begin{equation}
\varepsilon_\mathrm{AR}(t) = \varepsilon^{\mathrm{max},21}_{\rm AR}\frac{\mathrm{SN}(t)}{\mathrm{SN^{\mathrm{max},21}}} \;,
\label{eq:emerg_act_old}
\end{equation}
where $\varepsilon^{\mathrm{max},21}_{\rm AR}$ and $\mathrm{SN^{\mathrm{max},21}}$ are the three-month averaged emergence rate and SN value observed during the maximum of cycle 21 \citep[taken from][]{Schrijver-Harvey:1994}, respectively. 
Because at the time that the model was originally developed,  large-scale studies of magnetic flux emergence and evolution could not resolve internetwork fields, the earlier model was restricted to ERs as the only  magnetic bipoles smaller than ARs.
The emergence rate of the ERs, $\varepsilon_{\mathrm{ER}, n}$, of the cycle $n$ was not well known and was assumed to be a sinusoidal function, $g^n$, which was stretched such that the length of the ER cycle was longer than the respective sunspot cycle.
The amplitude of the ER cycle was simply taken to be proportional to the maximum value of the emergence rate of the ARs of the cycle, so that the emergence rate of ERs over a cycle had the form: 
\begin{equation}
\varepsilon_{\mathrm{ER},n}= \varepsilon^{\mathrm{max},n}_\mathrm{AR}Xg^n \;, 
\label{eq:emerg_eph_old}
\end{equation}
where
$X$ is an amplitude factor (a free parameter of the model; same for all cycles). 
Importantly, Eq.~\ref{eq:emerg_eph_old} implies a linear relationship between the emergence rate $\varepsilon_{\mathrm{ER},n}$ and the sunspot number. This, along with the linear relationship between $\varepsilon_{\mathrm{AR},n}$ and sunspot number (Eq.~\ref{eq:emerg_act_old}) results in an absence of flux emergence during extended periods of spotless days.

\begin{figure}[htb]
\centering
\resizebox{\hsize}{!}{
\includegraphics{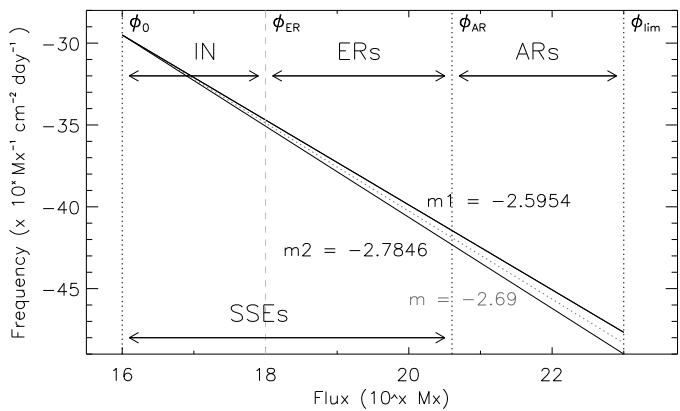}
}
\caption{Frequency of emergence vs. the unsigned flux of an emergence event (see Eq.~(\ref{eq:powerlaw}) following \citealp{Thornton-Parnell:2011}). $\phi_0$, $\phi_\mathrm{ER}$, $\phi_\mathrm{AR}$ and $\phi_\mathrm{limit}$ represent the limit below which the local dynamo flux dominates, the minimum ephemeral region flux, the minimum active region flux, and the upper limit of the active region flux, respectively.
The horizontal arrows mark the flux ranges corresponding to the internetwork (IN), ephemeral regions (ERs), active regions (ARs), and Small-Scale Emergences (SSEs). The SSE range includes IN fields and ERs.
The slope $m$ of the distribution was derived by \citet{Thornton-Parnell:2011} by fitting various observations at different activity levels. Slopes $m_1$ and $m_2$ represent the corresponding distributions at maximum and minimum of solar activity levels for cycle 21, respectively (see main text for details).
}
\label{fig:powerlaw}
\end{figure}


To overcome this shortcoming, we incorporate more recent solar observations by \citet{Thornton-Parnell:2011} into the model.
Using high-resolution Hinode, Solar Optical Telescope/Narrow-band Filter Imager (SOT/NFI) observations and combining them with earlier published data, they found that the emergence rate of the magnetic flux on the Sun follows a power-law distribution:
\begin{equation}
\frac{\mathrm{d} N}{\mathrm{d} \phi} = \frac{n_0}{\phi_0} \Big( \frac{\phi}{\phi_0} \Big)^{m} \;,
\label{eq:powerlaw}
\end{equation}
where $\phi_0=10^{16}$\,Mx (the smallest flux per feature that they include in their histograms),  $n_0=3.14\times10^{-14}$\,cm$^{-2}$\,day$^{-1}$ and  $m=-2.69$
(see the illustration in Fig.~\ref{fig:powerlaw}).

We note that \citet{Thornton-Parnell:2011} have summarised the results from multiple studies with a wide range of solar activity levels and observing conditions.
In earlier studies, \cite{Harvey-Zwaan:1993} and \citet{Harvey_PhD:1993} found that the emergence rate of ARs varied significantly more between solar activity minimum and maximum than that of ERs.
Whereas roughly 8.3 times more ARs emerged during the maximum of cycle 21 than during the minimum (the factor generally grows with the size of the regions, but was on average about 8.3 for all the ARs they studied), this ratio was roughly two for ERs. The number of the smallest magnetic features, forming the internetwork magnetic fields and having fluxes of $10^{16}$-- $10^{17}$\,Mx, appears to be nearly invariable over an activity cycle \citep{Buehler+al:2013,Lites:2014}. These features differ from the larger ones in that they are mainly brought about by a small-scale turbulent dynamo \citep{Voegler-Schuessler:2007,Rempel:2014} that produces the same amount of magnetic flux nearly independently of large-scale activity. 

To satisfy these observational constraints,  
on the one hand, we keep the number of the smallest magnetic features considered here (with a flux per feature of $10^{16}$\,Mx) fixed at all times. On the other hand, we allow the exponent $m$ to vary 
\citep[cf.][]{Parnell_2009,Schrijver-Harvey:1989}
around the empirical value $m = -2.69$ found by \citet{Thornton-Parnell:2011}
within the range 
$m_1 \geq  m \geq m_2$,
where $m_1 = m + \Delta m$ and $m_2 = m - \Delta m$. 
The slopes $m_1$ and $m_2$ describe the distributions of emergence rates during periods when the observed sunspot numbers are $\mathrm{SN}_1$ and $\mathrm{SN}_2$ (with SN$_1 > \mathrm{SN}_2$).
In our model, the slope $m$ follows the SN, $m(\mathrm{SN})$, 
according to the non-linear relationship:
\begin{equation}
m\left(\mathrm{SN}\right) = m_1 - \left(\mathrm{SN}_1^{\alpha} - \mathrm{SN}^{\alpha}\right) \frac{m_1-m_2}{\mathrm{SN}_1^{\alpha}-\mathrm{SN}_2^{\alpha}},
\label{eq:m}
\end{equation}
where $\alpha$ is a free parameter, fixed by comparing the output of the model to observations and independent reconstructions (see Sect.~\ref{sec:parameters}).

Now, the emergence rate of magnetic flux in ARs and SSEs at any given time can be calculated as:
\begin{equation}
\varepsilon_\mathrm{AR} =\int_{\phi_\mathrm{AR}}^{\phi_\mathrm{limit}} \frac{n_0}{\phi_0} \Big( \frac{\phi}{\phi_0} \Big)^m\;\phi\;\mathrm{d}\phi= \frac{n_0}{(m+2) \;{\phi_0}^{m+1}} \Big( \phi_\mathrm{limit}^{m+2} - \phi_\mathrm{AR}^{m+2} \Big) \;,
\label{eq:AR_emergence}
\end{equation}
and 
\begin{equation}
\varepsilon_\mathrm{SSE} = \int_{\phi_0}^{\phi_\mathrm{AR}} \frac{n_0}{\phi_0} \Big( \frac{\phi}{\phi_0} \Big)^m\;\phi\;\mathrm{d}\phi= \frac{n_0}{(m+2) \;{\phi_0}^{m+1}} \Big( \phi_\mathrm{AR}^{m+2} - \phi_0^{m+2} \Big) \;.
\label{eq:ER_emergence}
\end{equation}
Here $\phi_\mathrm{AR} =4\times10^{20}$\,Mx denotes the magnetic flux of the smallest bipolar regions hosting sunspots, i.e. the smallest active regions \citep[e.g.,][]{Zwaan:1978,Schrijver-Zwaan:2000,vanDriel:2015} and
$\phi_\mathrm{limit}$ is
the flux of the largest considered ARs. Since such regions are extremely rare,
the exact value of $\phi_\mathrm{limit}$ is not important.
Following \citet{Parnell_2009} and \citet{Thornton-Parnell:2011},
we take $\phi_\mathrm{limit}=10^{23}$\,Mx, which is somewhat larger than the maximum flux ($3\times 10^{22}$\,Mx) for ARs listed by \citet{Schrijver-Zwaan:2000} and \citet{vanDriel:2015}.
Tests have shown that also the exact value of $\phi_\mathrm{AR}$ adopted here plays only a minor role for the end result in the sense that although the free parameters may have slightly different values, the computed open and total magnetic flux remain almost unchanged.

To estimate how the slope $m$ of the distribution given by Eq.~(\ref{eq:powerlaw}) changes with activity (Eq.~(\ref{eq:m})) we rely on the observations by
\citet{Harvey-Zwaan:1993} and \citet{Harvey_PhD:1993} for cycle~21.
They found that the number of emerging ARs
in cycle 21
varied between the activity maximum and minimum by a factor of roughly 8.3.
The monthly-smoothed sunspot numbers corresponding to these periods (1979 -- 1982 for the maximum, as well as 1975 -- 1976 and 1985 -- 1986 for the preceding and following minima)
are then 
$\mathrm{SN}_1 =217$ and $\mathrm{SN}_2 = 17$, respectively.
These values are obtained for ISN2.0.
For GSN, these numbers correspond to $\mathrm{SN}_1 =130$ and $\mathrm{SN}_2 = 10$.
By using the factor of 8.3 found for the emergence frequency of ARs between activity maximum and minimum by \citet{Harvey_PhD:1993} as a constraint, we obtain $\Delta m = 0.0946$ (and thus $m_1 = -2.5954$ and $m_2 = -2.7846)$. 

\begin{figure}
\centering
\resizebox{\hsize}{!}{
\includegraphics{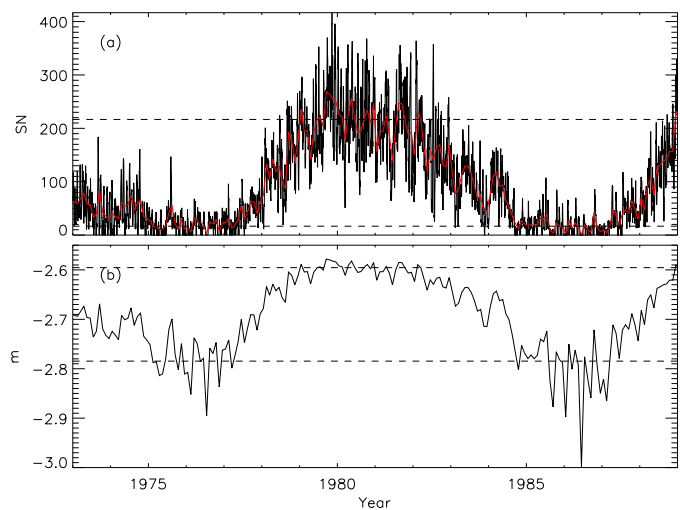}
}
\caption{
(a) Daily (black) and 30-day smoothed (red) sunspot number (ISN, v2.0) over cycle 21, including the adjacent mimima. The average values during the activity maximum (SN$_1$ = 217) and minimum (SN$_2$ = 17) of cycle 21 are marked by the horizontal dashed lines. (b) Evolution of the monthly-smoothed power-law slope $m$ over the same period, the horizontal dashed lines mark $m_1$ = -2.5954 corresponding to SN$_1$ = 217 and $m_2$ = -2.7846 corresponding to SN$_2$ = 17.}
\label{fig:SN_m_isn}
\end{figure}

Note that the values of $m$ reach values higher than $m_1$ or lower than $m_2$ at times when 
the sunspot number is higher than SN$_1$ or lower than SN$_2$, respectively. 
In particular, when the sunspot number is zero, the corresponding $m$ values are $m(0)=-3.952$ for ISN2.0 and $m(0)=-3.677$ for GSN.
The value of $m(0)$ is different for ISN2.0 and GSN because the value of $\alpha$, the free parameter of the model (see Table \ref{tbl:paras}, and Sect.~\ref{sec:parameters}) is different: $\alpha=0.059$ for ISN2.0 and $\alpha=0.075$ for GSN. 
The largest value of $m$
that the direct sunspot records give is obtained for the peak of cycle 19, with $\mathrm{max}(\mathrm{SN})=370$ (ISN2.0) and 247 (GSN), which results in $m(370)=-2.552$ (ISN2.0) and $m(247)=-2.542$ (GSN).
Figure \ref{fig:SN_m_isn}a shows the daily (black) and the monthly-smoothed (red) ISN2.0, while the evolution of $m_\mathrm{SN}$ computed from the monthly-smoothed ISN for the solar cycle~21 is shown in Fig.~ \ref{fig:SN_m_isn}b.

\subsection{Parameters of the model}
\label{sec:parameters}

\begin{table}[htb]
\centering
\caption{Parameters of the model. 
}
\begin{tabular}{l|cc|c}
     \hline
     Parameter   &  ISN  & GSN & W18 (ISN)\\
     \hline
$n_0$, cm$^{-2}$\,day$^{-1}$   & \multicolumn{2}{c|}{$3.14\times10^{-14}$} & \ldots\\
$\phi_0$, Mx & \multicolumn{2}{c|}{$10^{16}$} & \ldots\\
$\phi_\mathrm{AR}$, Mx & \multicolumn{2}{c|}{$4 \times 10^{20}$} & \ldots\\
$\phi_\mathrm{limit}$, Mx & \multicolumn{2}{c|}{$10^{23}$} & \ldots\\
SN$_1$ & 217 & 130 & \ldots\\
SN$_2$ &  17 &  10 & \ldots\\ 
$m$    & \multicolumn{2}{c|}{-2.69} & \ldots\\
$\Delta m$ & \multicolumn{2}{c|}{0.0946} & \ldots\\
$\tau^{\rm 0}_{\rm AR}$, yrs& \multicolumn{2}{c|}{$0.027$} & \ldots\\ 
$\tau^{\rm 0}_{\rm SSE}$, yrs& \multicolumn{2}{c|}{$1.1\times 10^{-5}$} & \ldots\\
     \hline   
$\alpha$ &  0.059 & 0.075 & \ldots\\
$\tau^{\rm r}_{\rm open}$, yrs& 0.16 & 0.09 & 0.14 \\
$\tau^{\rm s}_{\rm open}$, yrs& 3.98 & 3.90 & 3.75 \\
$\tau^{\rm r}_{\rm AR}$, yrs& 2.79 & 1.57 &  2.6 \\
$\tau^{\rm s}_{\rm AR}$, yrs& 88.15 & 89.81 & 88.3 \\
$\tau^{\rm s}_{\rm SSE}$, yrs& 10.15 & 10.11 & 20.6\tablefootmark{*}\\
     \hline
\end{tabular}
\label{tbl:paras}
\tablefoot{The upper part of the table lists parameters that are fixed, whereas the lower part lists free parameters. For comparison, the last column (W18) lists the values of the free parameters of the old model from the most recent version by \citet{Wu18_ssi}.\\
\tablefoottext{*} This value was obtained by \citet{Wu18_ssi} for ERs rather than SSEs.}
\end{table}

\begin{figure*}[!htb]
\sidecaption
\includegraphics[width=12cm]{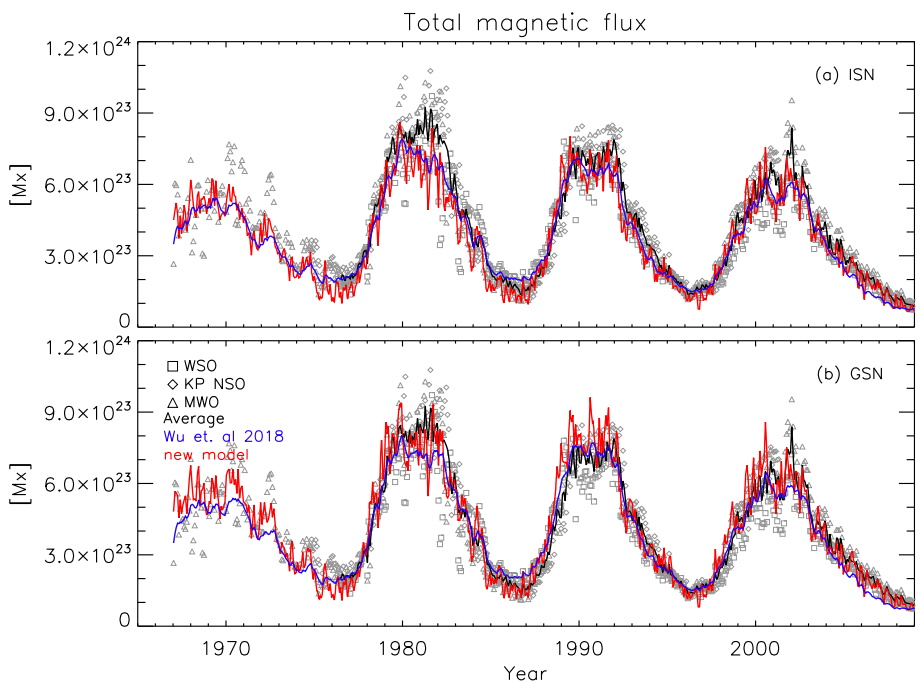}
\caption{Evolution of the total magnetic flux over cycles 21--24 computed with the new (red) and old (blue) models, as well as the observed flux. For observations, symbols show individual Carrington rotations and observatories, as indicated in the legend, and black solid line shows their average.
For the models, means of the daily total magnetic flux over the Carrington rotations are shown, and the contribution of the SSEs is reduced by a factor of 0.4 (see text for details).
Panels (a) and (b) use ISN and GSN as input, respectively.
}
\label{fig:flux_compare_total}
\end{figure*}

Our model, described in Sect.~\ref{sec:emerge}, has a number of parameters, summarised in Table~\ref{tbl:paras}.
The upper part of the table lists quantities taken or deduced from the literature and kept fixed throughout the analysis, whereas the five free parameters that are allowed to vary when fitting the data sets described in Sect.~\ref{sec:data} are given in the lower part.
One positive feature of the new model is that it has less free parameters than the old model (the old model required two free parameters to describe the emergence of ERs; see Table~1 of \citealp{Wu18_ssi}), giving it less `wriggle room' for reproducing observational data.  

All parameters relevant to the emergence rates of ARs and SSEs (Eqs.~\ref{eq:powerlaw}--\ref{eq:ER_emergence}) have been described in the previous section. Here we additionally comment on the decay and transfer times of the various components of the magnetic flux used in the ordinary differential equations describing the flux evolution (Eqs.~\ref{eq:ode1}--\ref{eq:ode6}).

The decay times of the ARs and SSEs, $\tau^{\rm 0}_{\rm AR}$ and $\tau^{\rm 0}_{\rm SSE}$, are estimated using the observations by \citet{Parnell_2009} and \citet{Thornton-Parnell:2011}.
Whereas \citet{Thornton-Parnell:2011} have analysed the emergence rate of different features as a function of their flux, \citet{Parnell_2009} analysed the magnetic flux for all (i.e. instantaneously) observed features.
By dividing the total number of the features of a given flux observed at a given instance by their emergence rate, we arrive at their mean lifetime.
The lifetime of the features increases with their sizes or fluxes.
For our purpose we make a simplification and calculate the lifetimes of ARs and SSEs as averages over all regions with fluxes above and below $\phi_\mathrm{AR}$, respectively, thus obtaining $\tau^{\rm 0}_{\rm AR}\approx 10$~days and $\tau^{\rm 0}_{\rm SSE}\approx 6$~minutes. 
Since the features with the smallest flux dominate the power law distribution, $\tau^{\rm 0}_{\rm AR}$ is short compared with lifetimes of large ARs and is closer to lifetimes of small ARs \citep[e.g., Table 1 of the review by][]{vanDriel:2015}. For the same reason the SSE lifetime is close to that of internetwork elements, rather than of ERs.
To compare better with observations of ERs, we therefore introduce $\phi_\mathrm{ER}=10^{18}$\,Mx (see Fig.~\ref{fig:powerlaw}), which denotes roughly the lowest magnetic flux contained within ERs.
If we now consider as ERs only the regions with $\phi_\mathrm{ER} < \phi < \phi_\mathrm{AR}$, then we obtain a lifetime $\tau^{\rm 0}_\mathrm{ER}\approx 2$~hours, which is comparable to the lifetimes of $\approx$hours to a day for regions with fluxes between $3\times 10^{18}$\,Mx and $10^{20}$\,Mx listed by \citet{vanDriel:2015}.
The maximum to minimum change of the flux emerging in ERs (i.e. with fluxes $\phi_\mathrm{ER} < \phi < \phi_\mathrm{AR}$) in our model is roughly a factor of 2.5, which is consistent with the results by \citet{Harvey_PhD:1993} and \citet{Harvey-Zwaan:1993}. 
We stress that this distinction into ERs and internetwork fields is only used for comparison purposes. Within the model they are not distinguished (see Sect.~\ref{sec:emerge}).

The decay timescales of the rapid, $\tau^{\rm r}_{\rm open}$, and slow, 
$\tau^{\rm s}_{\rm open}$,  open flux,
as well as timescales for the
flux transfer from ARs and SSEs to slow open flux, $\tau^{\rm s}_{\rm AR}$ and $\tau^{\rm s}_{\rm SSE}$, respectively, and the timescale for the transfer of the AR flux to the rapid open flux, $\tau^{\rm r}_{\rm AR}$, are free parameters of the model (together with $\alpha$ governing the change of the slope $m$ with the SN, as described in Sect.~\ref{sec:emerge}).
These parameters are fixed by comparing total and open magnetic flux with the corresponding observations listed in Sect.~\ref{sec:data}. 

To do this, we use the genetic algorithm PIKAIA \citep{Charbonneau:1995},
which searches for the set of parameters minimising the difference between the modelled and the reference data sets. 
We minimise the sum of the reduced chi-squared values, $\chi^2$, taking the errors of the observations and the number of data points into account. In other words, we search for the maximum of $1/(\chi^2_\mathrm{total}
+ \chi^2_\mathrm{OMF})$; see \citet{VieiraSolanki2010} for details and \citet{Dasi:2016} for a discussion of uncertainties in the parameter fitting.
The best-fit values of the parameters are listed in Table~\ref{tbl:paras}.

For comparison, the last column of Table~\ref{tbl:paras} also lists the values 
of the five free parameters that are already present in the VS2010 model, as obtained by \citet{Wu18_ssi} for the most recent version of the model. This version used the same ISN2.0 input record extended back with the data from \citet{Vaquero15} as done here.
The values of the parameters are very close in the two models, 
except $\tau^{\rm s}_{\rm SSE}$, for which \citet{Wu18_ssi} obtained 20.6~years compared to our 10.15~years (10.11~years when using GSN). 
Note, however, that \citet{Wu18_ssi} considered ERs rather than SSEs.
Interestingly, the values we obtain are close to the value of 10.08~years found to produce a best-fit by \citet{VieiraSolanki2010} and is within the range 10--90~years that is consistent with observation \citep[see][]{VieiraSolanki2010}.

The value of about 4~years for the decay time of the slow open flux $\tau^{\rm s}_{\rm open}$ is close to the radial decay term with a timescale of 5~years introduced into surface flux transport simulations by
\citet{Schrijver:2002} and \citet{Baumann:2006} to act on the unipolar fields at the solar poles. This was needed to reproduce the observed polar field reversals. Similarly, the $\tau^{\rm r}_{\rm open}$ obtained here (Table~\ref{tbl:paras}) is consistent with the estimate of the decay of closed flux carried by interplanetary coronal mass ejections, between 1 and 5 AU of between 38 and 55 days by \citet{Owens:2006}.

\begin{figure*}
\centering
\includegraphics[width=17cm]{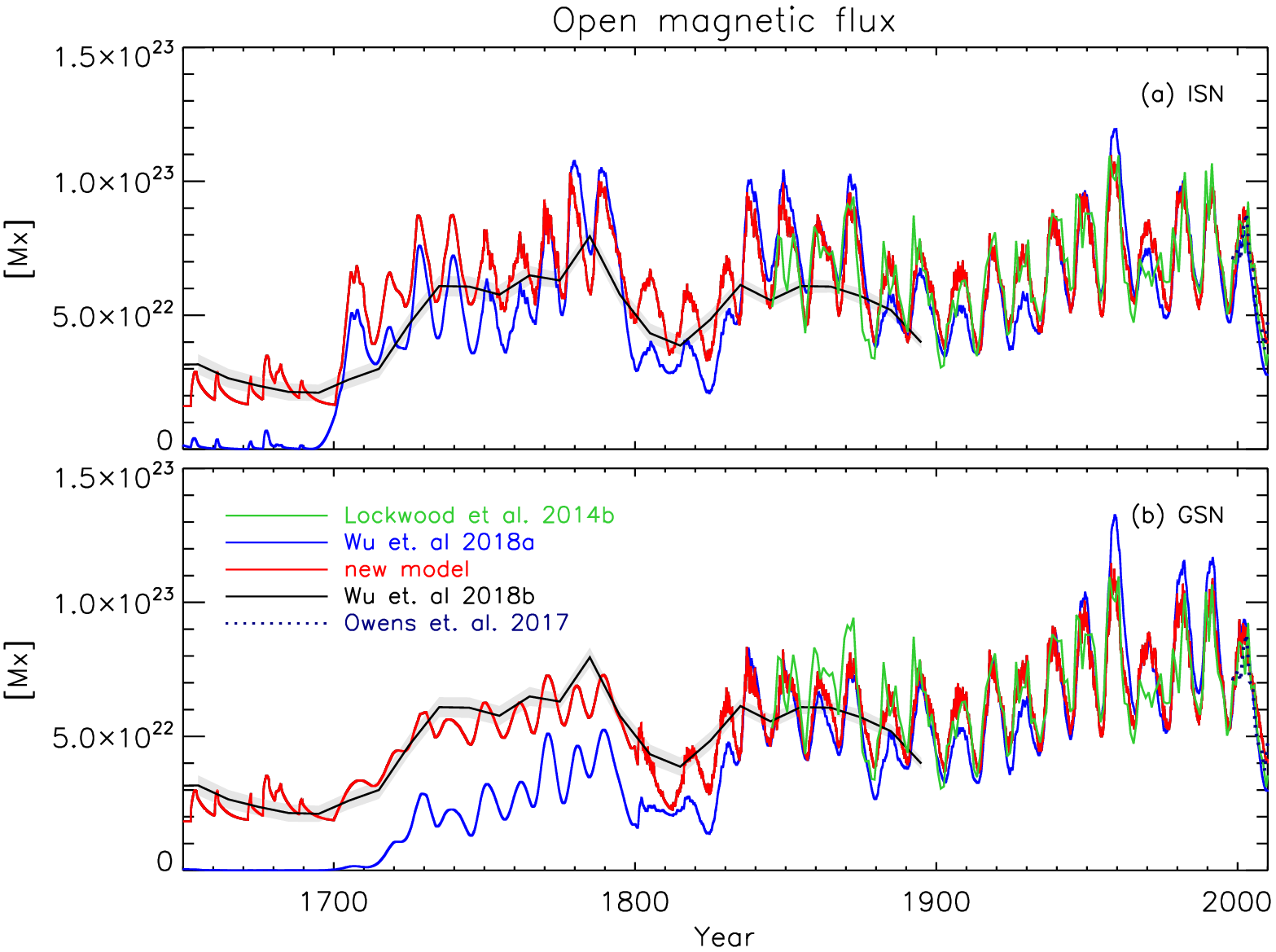}
\caption{
Open magnetic flux computed with the new (red) and old (blue) models, together with the empirical reconstruction from the geomagnetic aa-index \citep[][green]{Lockwood14_geo}.
Also shown are the reconstruction by \citet[][black solid line with shading marking the uncertainty]{Wu18_composite} from the decadal 
INTCAL13 $^{14}$C data \citep{reimer13} and the in-situ measurements from \citet[][dotted blue line]{Owens:2017} since 1998. Panels (a) and (b) use ISN and GSN as input, respectively.
}
\label{fig:flux_compare_open}
\end{figure*}

\section{Reconstruction of the total and open magnetic flux}
\label{sec:resu}

The computed total magnetic flux from 1965 onward is plotted in Fig.~\ref{fig:flux_compare_total}.
Following \citet{Krivova2007},
to account for the flux undetected due to the limited spatial resolution of observations \citep[see][]{Krivova:2004}, the contribution of the ER flux was multiplied by a factor of 0.4 before adding it to the contribution from the ARs and the open flux. 
Shown are means over the daily values for each Carrington rotation.
The new and old models are depicted in red and blue, respectively. Symbols show the observations: Wilcox Solar Observatory 
(WSO, squares), National Solar Observatory at Kitt Peak (NSO KP, diamonds), and Mount Wilson Observatory (MWO, triangles). Each individual symbol represents the total photospheric magnetic flux over a given Carrington rotation. 
To compute the $\chi^2$ value, we use the average of the three (1976--2002) or two (after 2002, see Sect.~\ref{sec:data}) datasets indicated by the black solid line.

The computed open magnetic flux is shown in Fig.~\ref{fig:flux_compare_open}.
The new model is shown in red, the old model is in blue,
and the reconstruction from the geomagnetic aa-index by \citet{Lockwood14_geo} is represented by the green line. Also shown are the in-situ measurements by \citet[][dotted blue line]{Owens:2017} since 1998 and a reconstruction (black solid line with shading indicating the uncertainty) of the open flux by \citet{Wu18_composite} from independent decadal INTCAL13 $^{14}$C data \citep{reimer13}.
Note that the underlying $^{14}$C data and thus also the OF reconstructed from them are decadal averages.
The agreement between our model and the $^{14}$C-based reconstruction is quite good. 
Particularly impressive is the agreement in the level of the open flux during the Maunder minimum, which is where we expect to see the biggest improvement relative to the old model. We emphasise that this $^{14}$C-based record was not used to constrain our model.
(Note that in the old model, the GSN record was used without the data from \citet{Vaquero15} over the Maunder minimum, and the computed open flux in the old model is therefore essentially flat at the zero level.)
Interestingly, over the 19th and the first half of the 18th centuries, the GSN-based reconstruction is closer to the isotope-based open flux than the reconstruction from the ISN, which lies somewhat higher.  

\begin{table*} 
\centering
\caption{Comparison of the modelled open and total magnetic fluxes to observations and independent reconstructions, quantified through their relative differences in means (in \%) and reduced $\chi^2$ values (listed in brackets).
}
\begin{tabular}{l|cc|cc}
     \hline
 Input &  \multicolumn{2}{c|}{ISN} & \multicolumn{2}{c}{GSN} \\
model version & old  & new & old & new\\
     \hline
$F_\mathrm{total}$ & $-$7.7 (0.037) & $-$9.9 (0.058) & $-$4.1 (0.037) & 0.8 (0.058) \\
$F_\mathrm{open, L14}$ &  $-$2.2 (0.297) & 0.4 (0.176) & $-$4.6 (0.389) & $-$1.6 (0.219) \\  
\hline\\[-3mm]
$F_\mathrm{open, Wu18b}$\tablefootmark{*} &  
$-$15.1 (1.718) & 6.1 (0.630) & $-$48.5 (1.674) & $-$9.6 (0.230) \\ 
$F_\mathrm{open, O17}$ &  $-$15.4 (0.655) & 3.8 (0.236) & $-$3.5 (0.495) & 5.8 (0.252) \\ 
     \hline
\end{tabular}
\label{tbl:chi2}
\tablefoot{The top part of the table lists the data sets that were used for parameter fitting (the average of the total magnetic flux measurements and the open flux reconstruction by \citealt{Lockwood14_geo}), while the bottom part lists independent data sets that were not used for the optimisation (OF reconstruction from  $^{14}$C data by \citealt{Wu18_composite} and the in-situ measurements by \citealt{Owens:2017}).\\
\tablefoottext{*} For decadally-averaged  reconstructions. As $^{14}$C data used for the reconstruction by \citet{Wu18_composite} are decadal averages, only decadally-averaged values of the OF could be reconstructed. Thus, to compute the corresponding $\chi^2$ values, our reconstructions were re-sampled, too.}
\end{table*}

A quantitative comparison of the total and open magnetic fluxes resulting from the old and the new models with the observations and independent reconstructions is presented in
Table~\ref{tbl:chi2}. The table lists the relative difference in means and the $\chi^2$ values (in brackets) between the models and the data.
For the total magnetic flux, the results are quite similar for both versions of the model.
In both cases, the mean modelled total magnetic flux is somewhat closer to the observations when the GSN is used as input.
The absolute difference in the means is slightly higher or lower for the new model if ISN2.0 or GSN are used, respectively. 
The new $\chi^2$ values are somewhat higher than in the old model.
This is, however, primarily due to fact that, by model design, the variability of the ER component on time scales shorter than the solar cycle in the old model was essentially smoothed out (see Eqs.~\ref{eq:emerg_act_old}--~\ref{eq:emerg_eph_old}) resulting in weaker short-term fluctuations than in the new model
(see Fig.~~\ref{fig:flux_compare_open}).
Thus, if we smooth the total flux from both models with a 3-months window before comparing, the $\chi^2$ values for the old model remain almost unchanged (0.036 for both ISN2.0 and GSN), while those for the new model decrease to 0.043 for ISN2.0 and 0.047 for GSN. In all cases, the $\chi^2$ values are quite low.

For the open flux, the new model provides a notably better fit than the old model to all three alternative datasets.
In all but one case, the absolute mean differences are significantly lower for the new model.
The only exception is the GSN-based reconstruction versus the in-situ data by \citet{Owens:2017}, for which the absolute mean difference is slightly lower for the old model.
However, the results are quite close for both versions of the model in this case. Note also that these data cover only a short recent period of time, over which the two models do not differ significantly.
The $\chi^2$ values are lower for the new model in all cases.
The fit is poorest for the reconstruction based on the decadal values of $^{14}$C.
Very recently, new $^{14}$C-based activity measures with annual resolution were published by \citet{brehm:2021}.
An application of our model to the dataset of \citet{brehm:2021} will be subject of a separate publication (Usoskin et al., submitted).

\section{Summary and discussion}
\label{sec:Summary}

We have revised the simple model describing  the evolution of the Sun's global total and open magnetic flux, originally proposed by \citet{Solanki2000,Solanki2002} and improved by \citet{VieiraSolanki2010}. The new version of the model takes into account the observation that fluxes of magnetic features follow a single power law, including internetwork fields, ERs and ARs
(\citealt{Parnell_2009}; cf. \citealt{Anusha:2017}). It also takes into account the fact that emergence rates of magnetic bipoles with fluxes between $10^{16}$~Mx and $10^{23}$~Mx, i.e. from the smallest ERs (and large internetwork features) to the largest ARs, also follow a power law according to the analysis by \cite{Thornton-Parnell:2011}. 

We assume that the difference in emergence rates between the maximum and the minimum of solar activity is adequately described by the varying power-law exponent, affecting magnetic features with fluxes $\phi > 10^{16}$\,Mx. Thus, for the smallest features
there is no change in emergence rate over the solar cycle, while the ratio of emergence rates during maximum to minimum increases steadily with increasing magnetic flux.
Thus, the number of emerging ARs varies between the solar maximum and minimum of cycle 21 by a factor of 8.3, while this factor is 2.5 for ERs and close to unity for internetwork fields. These values are consistent with the respective ratios found by  \citet{Harvey-Zwaan:1993} and \citet{Harvey_PhD:1993}, while the fact that the flux in internetwork fields hardly changes over the cycle is in agreement with the results obtained by, e.g., \citet{Buehler+al:2013} and \citet{Lites:2014}.

Using the sunspot number time series as input, the model returns time series of the open and total magnetic flux of the Sun. These resulting time series reproduce the open magnetic flux between 1845 and 2010 reconstructed by \cite{Lockwood14_geo} and total magnetic flux averaged over individual Carrington rotations obtained by various observatories between 1976 and 2009.

The main novel feature of the results of the model is that, in contrast to the earlier versions of the model, the output open magnetic flux does not drop to essentially zero during the Maunder minimum when almost no sunspots  were present for multiple decades, in agreement with open flux reconstructed from $^{14}$C data \citep[e.g.,][]{Wu18_composite}.
This significant improvement is a result of the model allowing for a non-zero  emergence of magnetic flux in small-scale bipolar features (encompassing ERs and internetwork fields) even during extended periods when sunspots were not present, e.g. during the Maunder minimum and other, similar grand minima \citep[e.g.,][]{Usoskin:2007}.

Even with this major update, the model still has some room for further improvements. 
Because it uses sunspot numbers as input (the only data of solar activity available prior to the middle of the 19th century), it cannot properly treat variations in solar activity that are not reflected in the number of sunspots. This is particularly evident during grand minima. During such times sunspots are only occasionally visible, whereas cosmogenic isotopes continue to display cyclic variations. This suggests that, in the context of the present model, the slope $m$ continues to vary in a cyclical manner and can go lower than the lowest value we have obtained ($m(0)= -3.952$ for ISN2.0 and $m(0)= -3.677$ for GSN).

In the old version of the model the ER emergence was constructed as a smooth, sinusoidal function.
In the new model, ER (and internetwork) emergence rate closely follows that of sunspots and has therefore the same temporal resolution as the input sunspot number series.
As a consequence, the new model does not feature temporal lags or shifts to the corresponding sunspot cycle, as in the old model.
However, it does account for the finding by \citet{Harvey:1994}
that small-scale features (in her study ERs) belonging to a given cycle start emerging at a relatively high rate well before the sunspot cycle starts.
The internetwork is independent of sunspot emergence and is, thus, not associated with a particular sunspot cycle. These regions are simply a result of the dynamo not having completely switched off even at times when there are no sunspots visible.
In this way, the overlap between neighbouring solar magnetic cycles is naturally introduced. This was the main feature of the original model of \citet{Solanki2000} responsible for the change in the level of the Sun's open magnetic flux from one solar minimum to another, first noticed by \citet{Lockwood1999}.
At the same time, the new model allows for the dynamo to continue working and produce activity cycles during grand minima, which are sufficiently strong to modulate cosmic rays (see, e.g., Fig.~\ref{fig:flux_compare_open} for the evolution of the open flux during the Maunder minimum) and hence influence the production of cosmogenic isotopes  \citep{Beer+al:1998,Fligge+al:1999,Usoskin2001}, but are too weak to produce more than occasional sunspots.

The higher level of the magnetic flux during periods of very low solar activity (e.g. during the Maunder and the Dalton minima, see Fig.~\ref{fig:flux_compare_open})
will presumably lead to a weaker secular variation of the total solar irradiance (TSI) and in particular the rise of the TSI since the Maunder minimum \citep[cf.,][]{Yeo:2020}, an important parameter for understanding the influence of solar irradiance variability on long-term climate trends \citep[e.g.,][]{Gray:2010,Solanki:2013}.  The influence of the revised magnetic flux time series on TSI will be the subject of a future investigation.


Another important application of this model will be in reconstructing total magnetic flux and sunspot numbers from production rates of cosmogenic isotopes. Such an application will require the model to be inverted, as has successfully been done with the older version of the model \citep[e.g.,][]{Usoskin2003,Usoskin2004,Solanki2004,usoskin_AA_16,Wu18_composite}. Such work is ongoing and will be the subject of a follow-on publication.

\begin{acknowledgements}
We thank Mathew Owens for the helpful comments and stimulating discussion.
NAK and SKS acknowledge support by the German Federal Ministry of Education and Research (Project No. 01LG1909C). SKS received support from the Ministry of Education of Korea through the BK21 plus program of the National Research Foundation. 
BH was supported by the International Max-Planck Research School (IMPRS) for Solar System Science at the University of Göttingen.
IU acknowledges a partial support from the Academy of Finland (Projects ESPERA no. 321882).
\end{acknowledgements}


\bibliographystyle{aa}

\end{document}